\documentclass[12pt,draft]{article}

\usepackage{amsthm,amssymb,amsmath,amscd,amstext,mathrsfs}

\newtheorem{theorem}{Theorem}[section]
\newtheorem{definition}{Definition}[section]
\newtheorem{lemma}{Lemma}[section]

\newtheorem{conjecture}[theorem]{Conjecture}
\newtheorem*{remark}{{\it Remark}}

\newcommand{\nc}{\newcommand}
\nc{\stack}[2]{{\begin{array}{c}
\scriptstyle #1 \\ \scriptstyle #2 \end{array}} }
\nc{\C}{{\mathbb C}}  
\nc{\R}{{\mathbb R}}  
\nc{\HH}{{\mathbb H}}
\nc{\Z}{{\mathbb Z}}
\nc{\N}{{\mathbb N}}
\nc{\dd}{{\rm d}}
\nc{\ee}{{\rm e}}

\begin{document}

\title{A proof of the Geroch--Horowitz--Penrose formulation\\ 
        of the strong cosmic censor conjecture\\
         motivated by computability theory}

\author{G\'abor Etesi\\
\small{{\it Department of Geometry, Mathematical Institute, Faculty of
Science,}}\\
\small{{\it Budapest University of Technology and Economics,}}\\
\small{{\it Egry J. u. 1, H \'ep., H-1111 Budapest, Hungary
\footnote{{\tt etesi@math.bme.hu}}}}}

\maketitle

\pagestyle{myheadings}
\markright{G. Etesi: Strong cosmic censorship and computability}

\thispagestyle{empty}

\begin{abstract}
In this paper we present a proof of a mathematical version of the strong
cosmic censor conjecture attributed to Geroch--Horowitz and Penrose but 
formulated explicitly by Wald. The proof is based on the existence of 
future-inextendible causal curves in causal pasts of events on the future 
Cauchy horizon in a non-globally hyperbolic space-time.

By examining explicit non-globally hyperbolic space-times we find that in 
case of several physically relevant solutions these future-inextendible 
curves have in fact infinite length. This way we recognize a close 
relationship between asymptotically flat or anti-de Sitter, physically 
relevant extendible space-times and the so-called Malament--Hogarth
space-times which play a central role in recent investigations in the
theory of ``gravitational computers''. This motivates us to exhibit a 
more sharp, more geometric formulation of the strong cosmic censor
conjecture, namely ``all physically relevant, asymptotically flat or
anti-de Sitter but non-globally hyperbolic space-times are 
Malament--Hogarth ones''.

Our observations may indicate a natural but hidden connection between 
the strong cosmic censorship scenario and the Church--Turing thesis revealing 
an unexpected conceptual depth beneath both conjectures.
\end{abstract}

\centerline{PACS numbers: 04.20.Dw, 89.20.Ff}
\centerline{Keywords: {\it Strong cosmic censorship; Malament--Hogarth 
space-times; Church--Turing thesis}}


\section{Introduction}
\label{one}


Certainly the deepest conceptual question of classical general
relativity theory is the so-called {\it cosmic censor conjecture} first
formulated by R. Penrose four decades ago \cite{pen1}. Roughly speaking 
the conjecture claims that predictability, probably one of the most 
fundamental concepts of classical physics, remains valid in the realm of 
classical general relativity i.e., all ``physically relevant'' space-times 
admit well-posed initial value formulation akin to other field theories. 
Meanwhile there has been a remarkable progress which culminated in a
general satisfactory solution of the problem of existence and behaviour of 
{\it short-time} solutions to the Einstein's constraint equations 
\cite{kla-rod-sze} the cosmic censor conjecture deals 
with the existence and properties of {\it long-time} solutions \cite{chru} and 
is still ``very much open'' as Penrose says in \cite{pen4}. One may then 
wonder what is the reason of this? Is the cosmic censor conjecture merely a 
technically more difficult question or is rather a conceptually deeper problem? 
On the contrary of its expected unified solution the cosmic censor
conjecture has rather split up into several rigorous or less rigorous
formulations, versions during the course of time. Therefore we can say
that nowadays there are several ``front lines'' where ``battles'' for settling 
or violating the cosmic censor conjecture are going on. Far from being 
complete we can mention the following results on the subjectmatter.

The so-called {\it weak cosmic censor conjecture}  in simple terms postulates:
\vspace{0.1in}

\noindent {\bf WCCC} {\it In a generic (i.e., stable), physically relevant 
(i.e., obeying some energy condition), asymptotically flat space-time 
singularities are hidden behind event horizons of black holes.}
\vspace{0.1in}

\noindent The weak version can be formulated rigorously as a Cauchy
problem for general relativity and the aim is to prove or disprove that 
for ``generic'' or ``stable'' (in some functional analytic sense) initial 
values at least, event horizons do form around singularities in an 
asymptotically flat space-time (where the notion of a black hole exists).

The first arguments in favour to this weak form came from studying the
stability of the Schwarzschild event horizon under simple, linear
perturbations of the metric. An early attempt to violate the weak
version was the following. As it is well-known, a static, electrically
charged black hole has only two parameters, namly its mass and charge.
However if its charge is too high compared with its mass, event horizon do
not occur hence the singularity could be visible by a distant observer.
Consequently we may try to overcharge a static black hole in order to
destroy its event horizon (we may argue in the same fashion in case of
rotating black holes). However this is impossible as it was pointed out
by Wald \cite{wal1} in 1974. Another, more general but still indirect,
argument for the validity of weak cosmic censorship is the so-called
Riemannian Penrose inequality \cite{pen2} proved by Bray \cite{bra} and
Huisken--Ilmanen \cite{hui-ilm} in 1997. As an important step, the
validity of the weak version in case of spherical collapse of a scalar 
field was established by Christodoulou \cite{chr1,chr2} in 1999.

The {\it strong cosmic censor conjecture} proposes more generally that all 
events have cause that is, there exist events chronologically preceding 
them and these events form a spacelike initial surface in any reasonable 
space-time. This also implies that singularities, except a possible initial 
``big bang'' singularity, are invisible for observers:
\vspace{0.1in}

\noindent {\bf SCCC} {\it A generic (i.e., stable), physically relevant 
(i.e., obeying some energy condition) space-time is globally hyperbolic.}
\vspace{0.1in}

\noindent Therefore this strong version also can be formulated in terms of
a Cauchy problem but in this case we want to prove the inextendibility of
maximal Cauchy developments of ``generic'' or ``stable'' (again in some 
functional analytic sense) initial data. Apparently this problem requires 
different techniques compared with the weak version.

Concerning the strong censorship we have partial important results, too.
On the one hand its validity was proved by Chru\'sciel--Isenberg--Moncrief
\cite{chr-ise-mon} and Ringstr\"om for certain Gowdy space-times (for a
recent survey cf. \cite{rin}) while by Chru\'sciel--Rendall
\cite{chr-ren} in 1995 in the case of spatially compact and locally
homogeneous space-times such as the Taub--NUT geometry. On the other hand 
one may also seek counterexamples to understand the meaning of ``generic'' in
both the weak and strong versions. Many authors (e.g. 
\cite{bri-hor, chr1, her-hor-mae,hor-she}) found hints in several 
physically relevant situations for the violation of the weak or strong 
versions. A thin class of Gowdy space-times \cite{chr-ise-mon, rin} also 
lacks global hyperbolicity.

We may however find a kind of ``compromise'' between the two extremal
approaches: seeking a general proof or hunting for particular
counterexamples. This is the following. As it is well-known, the strong   
version is false in its simplest intuitive form. This means that there are 
several physically relevant space-times what is more: basic solutions to 
the Einstein's equation which lack global hyperbolicity i.e., the maximal
Cauchy development of the corresponding initial data set is extendible.
The Taub--NUT space-time is extendible and in this case global
hyperbolicity fails in such a way that strong causality breaks down on
the future Cauchy horizon in any extension. The
Reissner--Nordstr\"om, Kerr, (universal covering of) anti-de 
Sitter space-times are also extendible but in these cases global
hyperbolicity is lost in a different way: from the future Cauchy horizons 
of their extensions a non-compact, infinite portion of their initial surfaces 
is observable. Therefore we have to indeed allow a collection of 
counterexamples consisting of apparently ``non-generic'' i.e., ``unstable'' 
space-times. Indeed, there are indirect hints that these extendible solutions 
are exceptional and atypical in some sense: small generic perturbations of 
them turn their Cauchy horizons into real curvature singularity thereby 
destroying extendibilty and saving strong cosmic censorship 
\cite{daf1, daf2, hol,ori1,poi-isr}. 

Since nowadays we do not know any other type of violation we may roughly 
formulate the strong version as follows due to Geroch--Horowitz 
\cite{ger-hor} and Penrose \cite{pen3} from 1979 but explicitly formulated 
by Wald \cite[305p]{wal2} (also cf. Theorem \ref{GHP} here):
\vspace{0.1in}

\noindent {\bf SCCC-GHP} {\it If a physically relevant (i.e., obeying 
some energy condition) space-time is not globally hyperbolic then its Cauchy
horizon looks like either that of the Taub--NUT or that of the 
Kerr space-time.}
\vspace{0.1in}

\noindent In this formulation the highly complex question of ``genericity'' 
or ``stability'' has been suppressed and incorporated into that of 
Taub--NUT-like \cite{hol} and Kerr-like space-times \cite{ale-ion-kla}. Since 
this version focuses only on the causal character of extendible space-times 
instead of their non-genericity, we may expect a proof of this form using 
causal theoretic methods only (instead of heavy functional analytic ones). 
One aim of this paper is to rigorously prove this version using ideas 
motivated by recent advances in an interdisciplinary field connecting 
computability and general relativity theory.

Recently there has been a remarkable interest in the physical foundations 
of computability theory and the Church--Turing thesis. It turned out 
that algorithm theory, previously considered as a very mathematical field,
has a deep link with basic concepts of physics. 

On the one hand nowadays we can see that our apparently pure mathematical 
notion of a Turing machine involves indirect preconceptions on space, time, 
motion, state and measurement. Hence it is reasonable to ask whether different 
choices of physical theories put for modeling these things have some effect 
on our notions of computability or not. At the recent stage of affairs it seems 
there are striking changes on the whole structure of complexity and even 
computability theory if we pass from classical physics to quantum or 
relativistic theories. Even certain variants of the Church--Turing thesis 
cease to be valid in some cases.

For instance taking {\it quantum mechanics} as our background
theory the famous Chaitin's omega number, a typical non-computable 
real number, becomes enumerable via an advanced quantum computer
\cite{cal-pav}. An adiabatic quantum algorithm exists to attack 
Hilbert's tenth problem \cite{kie1,kie2}. Chern--Simons {\it topological 
quantum field theory} can also be used to calculate the Jones polynomial of 
knots \cite{fre}. In the same fashion if we use 
{\it general relativity theory}, powerful ``gravitational computers'' can be 
constructed, also capable to break Turing's barrier \cite{wel}: Hogarth 
proposed a class of space-times in 1994, now called as 
{\it Malament--Hogarth space-times} allowing non-Turing computations 
\cite{hog1,hog2}. Hogarth' construction uses 
anti-de Sitter space-time which is also in the focal point of recent 
investigations in high energy physics. In the same vein, in 2001 the 
author and N\'emeti constructed another example by exploiting properties 
of the Kerr geometry \cite{ete-nem}. This space-time is also relevant as 
being the only candidate in general relativity for the final state of a 
collapsed, massive, slowly rotating star. A general introduction to the topic 
is Chapter 4 of Earman's book \cite{ear}. 

On the other hand it is conjectured that these generalized computational
methods are not only significant from a computational viewpoint but they 
are also in connection with our most fundamental physical concepts such 
as the standard model and string theory \cite{bla,sch}. A relation 
between computability and gravity also has been examined in 
\cite{acq-gas, ger-har}.

The natural question arises if the same is true for ``gravitational 
computers'' i.e., is there any pure physical characterization of 
Malament--Hogarth space-times? In our previous letter we 
tried to argue that these space-times also appear naturally in the strong
cosmic censorship scenario \cite{ete}. Namely we claimed that
space-times possessing powerful ``gravitational computers'' form the
unstable borderline separating the allowed and not-allowed space-times
by the strong cosmic censor (but these space-times are still considered
as ``physically relevant''). 

In this paper we try to push this analogy further and claim that using
a concept emerging from the theory of ``gravitational computers'' we can
prove the Geroch--Horowitz--Penrose form of the strong cosmic censor
conjecture. The proof uses standard causal set theory only with the 
simple but key observation that if a globally hyperbolic space-time is 
future-extendible then in the causal pasts of events on its future 
Cauchy horizon future-inextendible, non-spacelike curves appear. These curves 
also play a crucial role in the theory of non-Turing computers in general 
relativity: Malament--Hogarth space-times are exactly those for which the 
aforementioned curves exist, are timelike and complete. But checking 
case-by-case several physically relevant maximally extended examples lacking 
global hyperbolictiy like Kerr, Reissner--Nordstr\"om, (universal cover of) 
anti-de Sitter we find that these future-inextendible curves are 
indeed timelike and have infinite length. But on the contrary the Taub--NUT 
and certain extendible polarized Gowdy space-times with toroidal spatial 
topology lack this property: the corresponding inextendible curves are 
incomplete.

This motivates us to sharpen the Geroch--Horowitz--Penrose version of the 
strong cosmic censor conjecture recalled above like this (cf. 
Conjecture \ref{sejtes} here):
\vspace{0.1in}

\noindent {\bf SCCC-MH} {\it If a physically relevant
(i.e., obeying some energy condition), asymptotically flat or asymptotically
hyperbolic (i.e., anti-de Sitter) space-time is not globally hyperbolic then 
it is a Malament--Hogarth space-time.}
\vspace{0.1in}

\noindent Note that this formulation continues to avoid the question of
``genericity'' or ``stability''. We cannot prove or disprove this version but 
call attention that this formulation sheds some light onto a possible deep 
link between the cosmic censorship scenario and computability theory as 
follows. Consider the following physical reformulation of the Church--Turing 
thesis:\footnote{For the concept of an ``artificial 
computing system'' and of a ``Turing computable function'' in particular 
and for further details in general, we refer to \cite[Chapter 2]{ete-nem} and 
references therein.}
\vspace{0.1in}

\noindent {\bf Ph-ChT} {\it An artificial computing system based on a 
generic (i.e.,  stable), relevant (i.e., obeying some energy condition) 
classical physical system realizes Turing-computable functions.}
\vspace{0.1in}

\noindent Note that this formulation---in contrast to versions like 
\cite[Thesis 2 and 2']{ete-nem}---is a quite democratic one because it 
does not {\it a priori} excludes the existence of too powerful computational 
devices; it just says that they must in one or another way be unstable 
(which is apparently true for the various devices in 
\cite{cal-pav, kie1, kie2, ete-nem, hog1, hog2}). Accepting that all 
artificial computing systems based on classical physics can be modeled by 
``gravitational computers'' as will be argued in Section \ref{three} here as 
well as accepting {\bf SCCC-MH} we can see that {\bf SCCC} and {\bf Ph-ChT} 
are roughly equivalent hence involve the same depth. This might serve as an 
explanation for the permanent difficulty present in all 
approaches to the strong cosmic censor conjecture scenario.

The paper is organized as follows. In Section \ref{two} we prove 
rigorously the Geroch--Horowitz--Penrose version of the conjecture using 
methods mentioned above (cf. Theorem \ref{GHP} here). 

Then in Section \ref{three} we introduce all the notions 
required by computability theory. These are the concept of a {\it 
Malament--Hogarth space-time} (cf. Definition \ref{mh} here) and that of a 
{\it gravitational computer}. Then a case-by-case study of explicit examples 
helps us to select those non-globally hyperbolic space-times which also 
possess the Malament--Hogarth property. In this framework we can then 
offer a sharper Geroch--Horowitz--Penrose-type version of the strong cosmic 
censor conjecture which relates it with the aforementioned physical 
formulation of the Church--Turing thesis (cf. Conjecture \ref{sejtes} here). 

In Section \ref{four} we conclude our paper and speculate what has 
actually been proved. 


\section{The strong cosmic censor conjecture}
\label{two}


Basic references for this section are \cite{haw-ell,wal2} but we
will also frequently use \cite{bee-ehr-eas}. Let $(M,g)$ be a connected, 
four dimensional, smooth, time-oriented Lorentzian manifold-without-boundary 
i.e., a space-time. 

In this paper a class of space-times will be considered whose members $(M,g)$ 
can be carefully constructed as follows. Take a smooth, globally hyperbolic 
space-time $(D(S), g'\vert_{D(S)})$ which is supposed to be a solution to 
the coupled Einstein's equation
\[r-\frac{1}{2}sg'\vert_{D(S)}=8\pi T+\Lambda g'\vert_{D(S)}\]
with cosmological constant $\Lambda$ and matter content represented by a
stress energy-tensor $T$ obeying the {\it dominant energy condition}. Suppose 
that this matter field is {\it fundamental} in the 
sense that the associated Einstein's equation can be adjusted into the form 
of a quasilinear, diagonal, second order system of hyperbolic partial
differential equations. In this case $(D(S), g'\vert_{D(S)})$ admits a 
well-posed initial value formulation as follows. There exists an 
initial data set $(S,h,k)$ where $(S,h)$ is a smooth Riemannian 
three-manifold which is supposed to be complete and $k$ is a smooth 
$(0,2)$-type tensor field satisfying the usual constraint equations 
and this initial data set is related to the original space-time 
$(D(S), g'\vert_{D(S)})$ in the usual way \cite{haw-ell,wal2}: the 
unique maximal Cauchy development of $(S,h,k)$ is $(D(S), g'\vert_{D(S)})$ 
i.e. this is the largest space-time in which $S$ is a connected, spacelike 
Cauchy surface and $h$ and $k$ are the induced metric and extrinsic 
curvature of $S$ as embedded into $(D(S), g'\vert_{D(S)})$, respectively. 

Let $(M',g')$ be a (not necessarily unique) at least continuous 
extension of $(D(S), g'\vert_{D(S)})$ if exists. If $(D(S), 
g'\vert_{D(S)})$ is inextendible in this sense then we put 
$(M',g'):=(D(S), g'\vert_{D(S)})$. Thus $(D(S),g'\vert_{D(S)})\subseteqq 
(M',g')$ is a proper isometric embedding and $D(S)$ is open in $M$. Then 
$S$ is a {\it Cauchy surface} in $(M',g')$ which is {\it partial} if 
$(D(S),g'\vert_{D(S)})$ is not identical to $(M',g')$. We shall also 
suppose that $(M',g')$ is a maximal extension in the sense that if 
$(M,g)$ is another Lorentzian manifold of the same dimension such that 
$(M',g')\subseteqq (M,g)$ is a continuous isometric embedding then 
$(M',g')=(M,g)$. It is of course clear that for any given initial data 
set its unique maximal Cauchy development is maximal as a Cauchy 
development hence it is reasonable to demand its extension to be maximal 
as a continuous extension as well in order to get rid of very artificial 
extensions constructed by e.g. unphysical puncturing, etc. Thus 
throughout the text we shall suppose that given any initial data set 
$(S,h,k)$ as above whose maximal Cauchy development $(D(S), 
g\vert_{D(S)})$ satisfies an Einstein equation with fundamental matter 
obeying the dominant energy condition then its further extension 
$(M,g)$, if exists, is also maximal as an extension (and maybe satisfies 
an extended Einstein equation as well).

Finally we note that all causal or set-theoretical operations (i.e., 
$J^\pm (\cdot )$, $\subseteqq$, $\cap$, $\cup$, taking complement, 
closure, etc.) will be taken in the space $M$ with its standard manifold 
topology throughout the text.

\begin{lemma}
Let $(M,g)$ be a space-time as above with a strictly partial Cauchy
surface $S\subset M$ i.e., the corresponding maximal Cauchy development 
satisfies $(D(S), g\vert_{D(S)})\subsetneqq (M,g)$. 

Then for any $q\in M\setminus D(S)$ either strong causality is violated in 
$J^\pm (S)\cap J^\mp (q)$ or $J^\pm (S)\cap J^\mp (q)$ is non-compact 
(or both can happen).
\label{ketlehetoseg}
\end{lemma}

\noindent {\it Proof.} Fix a point $q\in M\setminus D(S)$. 
Then by definition there exists at least one non-spacelike curve in $M$ from 
$q$ such that: its image lies in $J^\mp (q)$, it is past/future-inextendible 
and having no intersection with $S$. Furthermore in any extension i.e., 
isometric embedding $(D(S), g\vert_{D(S)})\subset (M,g)$ if $S'\subset M$ is a 
spacelike submanifold such that $S\subseteqq S'$ then writing 
$h':=g\vert_{S'}$ we obtain an induced embedding $(S,h)\subseteqq 
(S',h')$ yielding that actually $S=S'$ taking into account that $(S,h)$ is a 
complete Riemannian manifold by assumption. This implies that the 
aforementioned curve is contained within $J^\pm (S)\cap J^\mp (q)\subset M$ 
i.e., this subset contains a non-spacelike curve of which at least one 
endpoint is missing. Hence referring to \cite[Lemma 8.2.1]{wal2} either strong 
causality is violated in the subset $J^\pm (S)\cap J^\mp (q)$ or it cannot be 
compact as claimed. $\Diamond$
\vspace{0.1in}

\noindent Our next goal is to find future inextendible non-spacelike
curves in the causal pasts of events ``on'' or ``beyond'' the future Cauchy 
horizon in non-globally hyperbolic space-times.\footnote{The issue of 
finding past-inextendible non-spacelike curves in the causal futures of
points chronologically preceding the globally hyperbolic regime is
similar hence we restrict our attention to the former case.} We will be  
frequently using various limit curve theorems from \cite{bee-ehr-eas, 
haw-ell} however a more compact approach also exists based on an improved 
theorem \cite[Theorem 3.1]{min} of the same kind.

\begin{lemma}
Let $(M,g)$ be a space-time as above with a strictly partial Cauchy 
surface $S\subset M$ i.e., the corresponding maximal
Cauchy development satisfies $(D(S), g\vert_{D(S)})\subsetneqq (M,g)$. 

If $q\in J^+(S)\cap (M\setminus D(S))$ then there exists a non-spacelike 
future-directed parameterized half-curve 
$\lambda :\R^+\rightarrow J^+(S)\cap J^-(q)$ or $\lambda: [0,a)
\rightarrow J^+(S)\cap J^-(q)$ with $a<+\infty$ such that 
$\lambda (0)\in S\cap J^-(q)$ and $\lambda$ has no future endpoint in 
$J^+(S)\cap J^-(q)$.
\label{vegnelkuligorbe}
\end{lemma}

\noindent {\it Proof.} By virtue of the previous lemma we can see that
either (i) $J^+(S)\cap J^-(q)$ fails to be strongly causal or (ii) 
$J^+(S)\cap J^-(q)$ is non-compact (or both cases can happen).

First assume case (i) is valid and let $x\in J^+(S)\cap J^-(q)$ be such 
that strong causality fails in this point. If $x$ is an interior point
then this failure means that there is a neighbourhood $x\in U\subset 
J^+(S)\cap J^-(q)$ and a countable collection 
$U=V_1\supset\dots\supset V_n\supset\dots$ of open sets 
satisfying $x\in V_n$ and corresponding non-spacelike parameterized 
half-curves $\lambda_n :[0,a_n]\rightarrow M$ or $\lambda_n:\R^+\rightarrow M$ 
such that $\lambda_n$ intersects $V_n$ more than once for all $n\in\N$. 
Without loss of generality we can suppose that 
$\lambda_n(0)=p\in S\cap J^-(q)$ for all $n\in\N$. For a given $n\in\N$ it 
can happen that $\lambda_n$ has a future endpoint $q_n\in M$; if not then put 
$q_n:=\emptyset$. Then if we remove these points from $M$ then we get a 
sequence $\{\lambda_n\}$ of non-spacelike future-inextendible curves with past 
accumulation point $\lambda_n(0)=p\in M$ for all $n\in\N$. Consequently via 
\cite[Proposition 3.31]{bee-ehr-eas} the sequence $\{\lambda_n\}$ has a 
limit curve (in the pointwise sense, cf. \cite[Definition 3.28]{bee-ehr-eas}) 
i.e., a future-inextendible non-spacelike parameterized curve
\[\lambda :\left[ 0\:,\:\sup\limits_na_n\right)\longrightarrow 
M\setminus\bigcup\limits_{n\in\N}\{ q_n\}\]
with $\lambda (0)=p$. It is clear that this non-spacelike curve 
considered as a curve in $M\supseteqq M\setminus\cup_{n\in\N}\{ q_n\}$ 
cannot have future endpoint in $M$. Furthermore since $\lambda (0)\in 
S\cap J^-(q)$ we can suppose that its image is contained within 
$J^+(S)\cap J^-(q)$. This is because if the non-spacelike $\lambda$ 
happens to exit $J^-(q)$ then it cannot return into it anymore contradicting 
the fact that it intersects each $V_n\ni x$ more than once.

Now assume $x\in\partial J^-(q)$. Then, since $(M,g)$ is open, there is
a point $q'\in M$ such that $q\in J^-(q')$ is an interior point
therefore $J^-(q)\subset J^-(q')$ and $x$ is an interior point of
$J^-(q')$. Then we can find a curve $\lambda$ as above for $q'$. Taking
into account that the image of $\lambda$ is contained within $J^+(S)\cap 
J^-(q')$ for all points $q'$ with $q\in J^-(q')$ we obtain that actually
$\lambda$ lies within $J^+(S)\cap J^-(q)$ as desired.

Secondly suppose (ii) is valid. Then there is at least one 
sequence $\{ q_n\}$ in $J^+(S)\cap J^-(q)$ such that no subsequence 
of it converges in $J^+(S)\cap J^-(q)$. Exploiting the completeness of $(S,h)$ 
we can pick this sequence such that $q\notin\{ q_n\}$ and there is a point 
$p\in D(S)$ satisfying $\cup_{n\in\N}\{ q_n\}\subset J^+(p)$. Let 
$\lambda_n :[0,a_n]\rightarrow J^+(p)\cap J^-(q)$ be a non-spacelike 
parametrized curve connecting $p$ with $q_n$. If we remove the points of 
$\{ q_n\}$ from $J^+(p)\cap J^-(q)$ then $\lambda_n$'s are future-inextendible 
non-spacelike half-curves in the punctured set with a past accumulation 
point $\lambda_n(0)=p$. Consequently again via 
\cite[Proposition 3.31]{bee-ehr-eas} the sequence $\{\lambda_n\}$ has a limit 
curve (in the pointwise sense, cf. \cite[Definition 3.28]{bee-ehr-eas}) i.e., 
a future-inextendible non-spacelike parameterized curve 
\[\lambda :\left[ 0\:,\:\sup\limits_na_n\right)\longrightarrow\left( 
J^+(p)\cap J^-(q)\right)\setminus\bigcup\limits_{n\in\N}\{ q_n\}\] 
with $\lambda (0)=p$. Taking into account that $\{ q_n\}$ has no convergent 
subsequence, $\lambda$ remains inextendible even in $J^+(p)\cap J^-(q)$ that 
is, has no future endpoint. Extending $\lambda$ in the past of $p$ back to 
$S$ or taking its intersection with $S$ and reparameterizing if necessary, 
we obtain a non-spacelike future-directed half-curve with $\lambda (0)\in 
S\cap J^-(q)$ but without future endpoint as claimed. $\Diamond$
\vspace{0.1in}
   
\noindent Our next step is to understand how strong causality or
compactness of intersections of causal sets break down if we exit the 
globally hyperbolic region of a space-time.

We need a technical tool namely a convenient topology on the set of 
non-spacelike parameterized curves as follows 
(cf. e.g. \cite[Chapter 3]{bee-ehr-eas} or \cite[206p]{wal2}). 
Let $p\in J^+(S)$ and consider the set $P(S,p)$ of continuous non-spacelike 
parameterized curves starting on $S$ and terminating at $p$. 
Let $K\subset\R^+$ be compact and $V\subseteqq M$ be open and define 
$O_{K,V}\subseteqq P(S,p)$ by
\[O_{K,V}:=\{\lambda\in P(S,p)\:\vert\:\lambda (K)\subset V\} .\] 
That is, $O_{K,V}$ consists of all non-spacelike 
parameterized curves from $S$ to $p$ whose portions $\lambda (K)$ for a 
fixed $K\subset\R^+$ lie entirely within a fixed $V\subseteqq M$. The {\it 
compact-open topology} on $P(S,p)$ is generated by the sets $\left\{ 
O_{K,V}\:\vert\:\mbox{$K\subset\R^+$ compact, $V\subseteqq M$ 
open}\right\}$. In particular if some $O\subseteqq P(S,p)$ has the form  
$O=\bigcup O_{K,V}$ 
then it belongs to this topology hence is open but not all open subsets are 
of this form.\footnote{Note that in general the 
$O_{K,V}$'s do {\it not} form a basis for the compact-open topology.} The 
notion of convergence in the compact-open topology is 
the following. The sequence $\{\lambda_n\}$ in $P(S,p)$ {\it converges to 
$\lambda\in P(S,p)$ in the compact-open topology} if for every open 
$\lambda\in O\subseteqq P(S,p)$ there exists an integer $n_O\in\N$ such that 
$\lambda_n\in O$ for all $n>n_O$. This implies that if 
$\lambda_n\rightarrow\lambda$ ($n\rightarrow +\infty$) in the compact-open 
topology then for all $K\subset\R^+$ compact and $V\subseteqq M$ open such 
that $\lambda (K)\subset V$ there exists an integer $n_{K,V}\in\N$ such that 
$\lambda_n(K)\subset V$ for all $n>n_{K,V}$.

The following proposition is taken from \cite{ete} where its proof was
sketched only hence we present a detailed proof here.
\begin{lemma}
Let $(M,g)$ be a space-time as above. Let $S$ be a (partial) Cauchy
surface in it with maximal Cauchy development $(D(S),
g\vert_{D(S)})\subseteqq (M,g)$. If $x\in J^+(S)$ such that 
$\overline{S\cap J^-(x)}$ is compact and $J^+(S)\cap J^-(x)$ is strongly 
causal then $J^+(S)\cap J^-(x)$ is compact.
\label{kompaktsag}
\end{lemma}

\noindent {\it Proof.} Let $\{ q_i\}$ be an arbitrary sequence of points 
in $J^+(S)\cap J^-(x)$. We have to find a subsequence of it converging to a
point in $J^+(S)\cap J^-(x)$. If there was a subsequence $\{ q_j\}
\subset\{ q_i\}$ such that $q_j\rightarrow x$ ($j\rightarrow +\infty$) 
then we could finish the proof right now. Assume this is not the case. 
Then there is a neighbourhood $x\in U\subset M$ such that $U\cap 
J^+(S)\cap J^-(x)$ does not contain any point of $\{ q_i\}$.

First suppose that all but finite points of $\{ q_i\}$ sit in the
interior of $J^+(S)\cap J^-(x)$. In this case by exploiting (the first and 
last time here) the starting assumption that $(M,g)$ is maximal as a 
continuous extension hence is not punctured, etc. we can choose 
timelike future-directed parameterized half-curves 
$\lambda_i: [0,a_i]\rightarrow J^+(S)\cap J^-(x)$ such that 
\begin{enumerate}

\item[(i)] $\lambda_i(a_i)=x$ for all $i\in\N$ that is, all curves 
terminate at $x$;

\item[(ii)] there is a $t_i\in [0, a_i)$ such that $\lambda_i(t_i)
=q_i$ for all $i$ that is $\lambda_i$ intersects the point $q_i$ of the
sequence $\{ q_i\}$;

\item[(iii)] if $t_i>0$ then $\lambda_i(0)=:p_i\in S\cap J^-(x)$ for all
$i$ that is $\lambda_i$ departs from $S$ giving rise to a sequence $\{p_i\}$ 
in $S\cap J^-(x)$ (if $t_i=0$ for some $i$ then $p_i:=q_i$ in this case).
\end{enumerate} 

\noindent Note that the construction of this curves is highly non-unique. 
Nevertheless (i) and (iii) imply that $\{\lambda_i\}$ is a sequence in 
$P(S,x)$. Taking into account that $\overline{S\cap J^-(x)}$ is compact, there 
exists a subsequence $\{ p_j\}$ of the induced sequence $\{ p_i\}$ 
given by (iii) converging to a point $p\in\overline{S\cap J^-(x)}\subseteqq S$ 
since $S$ is closed in $M$. Consequently the 
corresponding subsequence $\{\lambda_j\}$ of our curves has a past 
accumulation point: $\lambda_j(0)=p_j\rightarrow p\in S$ 
($j\rightarrow +\infty$). 
Moreover if we remove $x$ from $J^+(S)\cap J^-(x)$, then $\{ p_j\}$ gives 
rise to a sequence of future-inextendible non-spacelike half-curves 
$\{\lambda_j\}$ by property (i) and (iii). Hence again by 
\cite[Proposition 3.31]{bee-ehr-eas} there exists a future-directed 
non-spacelike parameterized limit curve (again in the pointwise sense, cf. 
\cite[Definition 3.28]{bee-ehr-eas})
\[\lambda : \left[ 0\:,\: \sup\limits_ja_j\right)\longrightarrow \left( 
J^+(S)\cap J^-(x)\right)\setminus\{ x\}\] 
satisfying $\lambda (0)=p\in S$. Moreover $\lambda_j(a_j)=x$ for all 
$j\in\N$ hence $x$ is a future accumulation point of $\{\lambda_j\}$ 
thus in fact $a:=\sup_ja_j<+\infty$ and $\lambda (a)=x$. Thus 
$\lambda\in P(S,x)$. Therefore after reparametrizing we obtain a sequence 
$\{\lambda_j\}$ in $P(S,x)$ such that 
\[\left\{\begin{array}{ll}
\mbox{$\lambda_j: [0,a]\longrightarrow J^+(S)\cap J^-(x)$ 
for all $j\in\N$}\\
\mbox{$\lambda_j(0)\longrightarrow p$ ($j\rightarrow +\infty$)}\\
\mbox{$\lambda_j(a)\longrightarrow x$ ($j\rightarrow +\infty$)}
\end{array}\right.\] 
i.e., both endpoints of $\{\lambda_j\}$ 
converge. Since $J^+(S)\cap J^-(x)$ is strongly causal by assumption, there is 
a subsequence $\{\lambda_k\}\subseteqq\{\lambda_j\}$ such that 
$\lambda_k\rightarrow\lambda$ ($k\rightarrow +\infty$) in the compact-open 
topology on $P(S,x)$, too (cf. \cite[Proposition 3.34]{bee-ehr-eas} as 
well as \cite[Theorem 3.1]{min}). Thus for any $0\leqq\varepsilon$ and 
$V\subseteqq M$ open such that $\lambda ([\varepsilon,a-\varepsilon])\subset 
V$ one finds that $\lambda_k([\varepsilon,a-\varepsilon])\subset V$ for all 
$k>k_{\varepsilon, V}$. But $\lambda$ is a continuous image of the compact 
interval $[\varepsilon,a-\varepsilon]\subset\R^+$ 
consequently $\lambda ([\varepsilon,a-\varepsilon])\subset M$ is compact 
hence we can choose $V\subseteqq M$ and $0<\varepsilon$ so that $\overline{V}$ 
is compact and $\overline{V}\subset J^+(S)\cap J^-(x)$ holds.\footnote{Recall 
that the topology of $M$ is generated by the complete metric space $(M,d_h)$ 
associated to some auxiliary complete Riemannian metric $h$ put onto $M$ thus 
we can take $V$ to be an open ``narrow sausage'' surrounding 
$\lambda ([\varepsilon,a-\varepsilon])\Subset J^+(S)\cap J^-(x)$.} 
Since we already know that $p_k\rightarrow p\in J^+(S)\cap J^-(x)$ i.e. 
converges let assume that except finitely many cases $q_k\not=p_k$. Then 
$q_k\in\lambda_k ([\varepsilon,a-\varepsilon])\subset\overline{V}$ 
via property (ii) therefore by exploiting the compactness of $\overline{V}$ 
there is a convergent subsequence $\{ q_l\}\subseteqq\{q_k\}\subset
\overline{V}\subset J^+(S)\cap J^-(x)$. Thus eventually we succeeded to find 
a subsequence of the original arbitrary sequence $\{ q_i\}$ 
which is convergent in $J^+(S)\cap J^-(x)$ demonstrating the compactness of 
$J^+(S)\cap J^-(x)$.

Finally, if the points of $\{ q_i\}$ lie on $J^+(S)\cap\partial J^-(x)$
then take an $x'\in M$ such that $x\in J^-(x')$ is an interior point.
Then we have $J^-(x)\subset J^-(x')$ and the members of $\{ q_i\}$ are
interior points of $J^+(S)\cap J^-(x')$. Repeating the previous
procedure we obtain that $\{ q_i\}$ possesses a convergent subsequence
in $J^+(S)\cap J^-(x')$. But taking into account that this is true for
all $x'\in M$ with $x\in J^-(x')$ we find that in fact the accumulation
point of $\{ q_i\}$ is contained within $J^+(S)\cap J^-(x)$ as
claimed. $\Diamond$         
\vspace{0.1in}

\noindent After these preliminaries we are in a position to prove a variant of 
the strong cosmic censor conjecture attributed to Geroch--Horowitz 
\cite{ger-hor} and Penrose \cite{pen3} but formulated explicitly by Wald 
\cite[305p]{wal2}). Recall that $H^+(S):=\partial\overline{D^+(S)}$ is 
called the {\it future Cauchy horizon} of $D(S)\subset M$.

\begin{theorem} {\rm (the Geroch--Horowitz--Penrose version of the
strong cosmic censor conjecture)} Let $(S,h,k)$ be an intial data set
for Einstein's equation with $(S,h)$ a complete Riemannian
three-manifold and with a fundamental matter represented by a
stress-energy tensor $T$ obeying the dominant energy condition. 

Then, if the maximal Cauchy development of this initial data set is
extendible, for each $x\in H^+(S)$ in any extension\footnote{Cf. again our 
careful definition of an extension at the beginning of Section \ref{two}.} 
either strong causality is violated at $x$ or $\overline{S\cap J^-(x)}$ is 
non-compact.
\label{GHP}
\end{theorem}
\begin{remark}\rm Note that $\overline{S\cap J^-(x)}=\overline{S\cap I^-(x)}$ 
hence our statement coincides with \cite[305p]{wal2}. The theorem also 
implies that these extensions of course cannot be globally hyperbolic.
\end{remark}

\noindent {\it Proof.} Let $(D(S), g\vert_{D(S)})$ be the unique maximal Cauchy
development of $(S,h,k)$ and assume it admits a further at least 
continuos maximal extension $(M,g)$. Let $x\in H^+(S)$ be any point on the 
future Cauchy horizon in this extension. Then, by Lemma \ref{ketlehetoseg} 
the set $J^+(S)\cap J^-(x)$ cannot be strongly causal and compact. If strong 
causality is violated in this set, we get the first possibility of the theorem. 
Indeed, strong causality can fail only in $(J^+(S)\cap J^-(x))\cap H^+(S)$ 
because the remaining portion is globally hyperbolic. But we can repeat 
the previous procedure for $(J^+(S)\cap J^-(x))\cap U$ where $U$ is an 
arbitrary open set in $M$ containing $x$ yielding that this point must 
coincide with $x$ itself.

If strong causality is valid within $J^+(S)\cap J^-(x)$ then 
$\overline{S\cap J^-(x)}$ cannot be compact because in this case 
$J^+(S)\cap J^-(x)$ would be compact, too via Lemma \ref{kompaktsag} 
contradicting again Lemma \ref{ketlehetoseg}. Hence one obtains the second 
possibility of the theorem. $\Diamond$
\vspace{0.1in}

\noindent For a future comparison we give a refomulation of the above 
theorem based on Lemma \ref{vegnelkuligorbe}.

\begin{theorem} {\rm (reformulation of the Geroch--Horowitz--Penrose
version)}. Let $(S,h,k)$ be an initial data set as in Theorem \ref{GHP}.
Then, if the maximal Cauchy development of this initial data is
extendible, for each $x\in H^+(S)$ in any extension, $J^+(S)\cap J^-(x)$
contains a future-directed non-spacelike curve without future endpoint.
$\Diamond$
\label{GHP'}
\end{theorem}

\begin{remark}\rm 1. Notice that the Geroch--Horowitz--Penrose form of 
the strong cosmic censor conjecture deals with causal or conformal 
properties of a space-time only. This is also reflected in the 
mathematical structure of the proof: we were not forced to use hard 
analytical techniques to achieve the result. But note again that we assumed 
that our are extensions are maximal and satisfy an Einstein equation at 
least inside their globally hyperbolic domain consequently are free of 
removable singularities arising for instance from artificial 
puncturation of space-time, etc. 

2. However in fact we had to use the validity of the Einstein equation 
with some fundamental matter obeying the dominant energy condition only 
in an auxiliary way in the proof: it was only necessary to formulate a 
space-time as the unique solution to a Cauchy problem (implying this 
globally hyperbolic region being free of removable 
singularitites). Hence our results remain valid for a vast class of 
space-times which are still subject to the formulation as a Cauchy 
problem but satisfy more general field equations than Einstein's 
equation \cite{rac1, rac2}.

3. Of course we also have to pay some price for this approach: although we have 
been able to conclude that any extension has the desired causal property, we 
actually do not know whether or not these extensions are generic or 
unstable in any sense. Indeed, if the counterexamples mentioned in 
Section 1 turn out to be generic in some strict mathematical sense then 
extendibility of space-times with the Geroch--Horowitz--Penrose property must 
be generic, too that is, a stable phenomenon.
\end{remark}


\section{Malament--Hogarth space-times}
\label{three}


In this section we turn the coin and introduce the concept of a 
Malament--Hogarth space-time and that of a ``gravitational computer''. As a 
motivation we mention that in these space-times, at least in principle, one can 
construct powerful computational devices capable for computations beyond the 
Turing barrier. A typical example for such a computation is checking the 
consistency of ZFC set theory \cite{ete-nem, wel}. 

Let us consider the following class of space-times (cf. \cite{ear, 
ete-nem,hog1, hog2, wel}):

\begin{definition} Let $(S,h,k)$ be an initial data set for Einstein's
equation, with $(S,h)$ a complete Riemannian manifold. Suppose a
fundamental matter field is given represented by its stress-energy 
tensor $T$ satisfying the dominant energy condition. Let $(M,g)$ be a
maximal continuous extension (if exists) of the unique maximal Cauchy
development $(D(S),g\vert_{D(S)})$ of the above initial data set.
\begin{itemize}

\item[{\rm (i)}] Then $(M,g)$ is called a {\em Malament--Hogarth space-time} 
if there is a future-directed timelike half-curve 
$\gamma_C :\R^+\rightarrow M$ such that $\Vert\gamma_C\Vert =+\infty$ and 
there is a point $q\in M$ satisfying $\gamma_C (\R^+)\subset J^-(q)$. The 
event $q\in M$ is called a {\em Malament--Hogarth event}; 

\item[{\rm (ii)}] $(M,g)$ is called a {\em generalized Malament--Hogarth 
space-time} if there is a future-directed timelike half-curve 
$\gamma_C :\R^+\rightarrow M$ without future endpoint and there is a point 
$q\in M$ satisfying $\gamma_C (\R^+)\subset J^-(q)$. The event $q\in M$ is 
called a {\em generalized Malament--Hogarth event}.
\end{itemize}
\label{mh}
\end{definition}

\begin{remark}\rm 1. If $(M,g)$ is a (generalized) Malament--Hogarth 
space-time then there exists a future-directed timelike curve 
$\gamma_O :[a,b]\rightarrow M$ joining $p\in J^-(q)$ with $q$ satisfying 
$\Vert\gamma_O\Vert<+\infty$. The point $p\in M$ can be chosen to lie in the 
causal future of the past endpoint of $\gamma_C$. 

2. Moreover the reason we require fundamental matter fields obeying the
dominant energy condition, geodesically complete initial surfaces, extensions 
to be maximal etc., is that we want to exclude the very artificial 
examples of Malament--Hogarth space-times.

3. It follows from Theorem \ref{GHP'} here that all non-globally 
hyperbolic space-times are generalized Malament--Hogarth ones. But a 
sufficiently nice non-globally hyperbolic space-time is in fact 
{\it conformally equivalent} to a Malament--Hogarth-like space-time which 
is however probably not the solution of the Einstein's equation with a 
physically relevant matter content; indeed, any so-called distinguishable 
space-time can be conformally rescaled to be timelike and null geodesically 
complete (cf. \cite[Theorem 6.5]{bee-ehr-eas}). 
\end{remark} 

\noindent The motivation is the following (for details we refer to 
\cite{ete-nem}). Take any Malament--Hogarth space-time $(M,g)$. 
Consider a Turing machine realized by a physical computer $C$ moving along
the curve $\gamma_C$ of {\it infinite} proper time. Hence the physical
computer (identified with $\gamma_C$) can perform arbitrarily long
calculations in the ordinary sense. In addition there exists an 
observer $O$ following the curve $\gamma_O$ (hence denoted by $\gamma_O)$ of 
{\it finite} length such that he hits the Malament--Hogarth event $q\in 
M$ in {\it finite} proper time. But by definition 
$\gamma_C(\R^+)\subset J^-(q)$ therefore in $q$ he can receive the answer 
for a {\it yes or no question} as the result of an {\it arbitrarily long} 
calculation carried out by the physical computer $\gamma_C$. This is 
because $\gamma_C$ can send a light beam at arbitrarily 
late proper time to $\gamma_O$. Clearly the pair $(\gamma_C, \gamma_O)$ 
in $(M,g)$ with a Malament--Hogarth event $q$ is an artificial computing 
system i.e., a generalized computer in the sense of \cite{ete-nem}. 

Imagine the following exciting situation as an example. $\gamma_C$ is 
asked to check all theorems of our usual set theory (ZFC) in order to check    
consistency of mathematics. This task can be carried out by $\gamma_C$   
since its world line has infinite proper time. If $\gamma_C$ finds a
contradiction, it can send a message (for example an appropriately coded 
light beam) to $\gamma_O$. Hence if $\gamma_O$ receives a signal from 
$\gamma_C$ {\it before} the Malament--Hogarth event $q\in M$ he can be 
sure that ZFC set theory is not consistent. On the other hand, if $\gamma_O$  
does not receive any signal before $q$ then, {\it after} $q$, $\gamma_O$ can 
conclude that ZFC set theory is consistent. Note that $\gamma_O$ having 
finite proper time between the events $\gamma_O(a)=p$ (departure for the 
experiment) and $\gamma_O(b)=q$ (hitting the Malament--Hogarth event), he 
can be sure about the consistency of ZFC set theory within finite (possibly 
very short) time. This shows that certain very general formulations of the
Church--Turing thesis (for instance \cite[Thesis 2,2' and 3]{ete-nem}) 
cannot be valid in the framework of classical general relativity.

In general---keeping in mind the definition of a Malament--Hogarth
space-time---a quintuple $(M,g,q,\gamma_C, \gamma_O)$ is called a {\it
gravitational computer} if $(M,g)$ is a space-time, $\gamma_C$, $\gamma_O$
are timelike curves and $q\in M$ is an event such
that the curves lie within $J^-(q)$. This concept is broad enough to serve as
an abstract model for all kind of artificial computing systems based on 
classical physics so that an artificial computing system can perform 
non-Turing computations if and only if the corresponding gravitational 
computer is defined in an ambient space-time possessing the 
Malament--Hogarth property. Indeed, in the case of modeling a usual 
(Turing) artificial computing system the ambient space-time $(M,g)$ can be 
simply taken to be the Minkowskian or Newtonian one with any event $q\in M$ 
and curves $\gamma_C=\gamma_O$ in its causal past.\footnote{That is, the 
computer and the observer ``stay together'' during the course of the 
computation along a common worldline $\gamma_{CO}$ in $(M,g)$ which is 
moreover of finite length in practice.} 
However if the artificial computing system is expected to perform 
non-Turing computations then it is equivalent (cf. \cite[Chapter 
2]{ete-nem}) to a usual (Turing) artificial computing system with the only 
extra property of being able to solve at least once the so-called 
halting problem; this can be carried out if the ambient space-time $(M,g)$ is 
a Malament--Hogarth one with Malement--Hogarth event $q\in M$ and curves 
$\gamma_C$ and $\gamma_O$ as above.

One can raise the question if Malament--Hogarth space-times are
relevant or not from a physical viewpoint. We put off this very important
question for a few moments; instead we prove basic properties of
Malament--Hogarth space-times by evoking \cite[Lemmata 4.1 and 4.3]{ear}. 
These properties are also helpful in seeking realistic examples. 
\begin{lemma}
Let $(M,g)$ be a Malament--Hogarth space-time as above. Then $(M,g)$ is not
globally hyperbolic. Moreover, if $q\in M$ is a Malament--Hogarth event
and $S\subset M$ is a connected spacelike hypersurface such that 
$\gamma_C(\R^+)\subset J^+(S)\cap J^-(q)$ then $q$ is on or beyond the 
future Cauchy horizon $H^+(S)$ of $S$.
\label{phelye}
\end{lemma}

\noindent {\it Proof.} Consider a point $p\in M$ such that $\gamma_C (0)=p$. 
If $(M,g)$ was globally hyperbolic then $(M,g)$ would be strongly causal 
and in particular $J^+(p)\cap J^-(q)\subset M$ compact. 
We know that $\gamma_C(\R^+)\subset J^-(q)$ hence in fact 
$\gamma_C(\R^+)\subset J^+(p)\cap J^-(q)$. Consequently its future
(and of course, past) endpoint are contained in $J^+(p)\cap J^-(q)$ (cf. 
\cite[Lemma 8.2.1]{wal2}). However $\gamma_C$ is a causal curve with 
$\Vert \gamma_C\Vert =+\infty$ hence it is future
inextendible i.e., has no future endpoint. But this is impossible hence
$J^+(p)\cap J^-(q)$ cannot be compact or strong causality must be violated
within this set leading us to a contradiction.

Secondly, assume $q\in \overline{D^+(S)}\setminus\partial\overline{D^+(S)}$ 
i.e., $q$ is an interior point of the future domain of dependence of $S$. 
Then there is an $r\in D^+(S)$ chronologically preceded by $q$ (with respect
to some time function assigned to the Cauchy foliation of $D^+(S)$). 
Letting $N:=J^+(S)\cap J^-(r)$ then $N\subset D^+(S)$ hence $(N, 
g\vert_N)$ is a globally hyperbolic space-time containing the 
Malament--Hogarth event $q$ and the curve $\gamma_C$. Consequently we can 
proceed as above to arrive at a contradiction again. $\Diamond$
\vspace{0.1in}

\noindent By the aid of this one can provide a
characterization of Malament--Hogarth space-times \cite{ete}.

\begin{lemma}
Let $(M,g)$ be a Malament--Hogarth space-time with $q\in
H^+(S)\subset M$ a Mal\-am\-ent--Hogarth event. Consider a timelike
curve $\gamma_C$ as above with $\gamma_C(\R^+)\subset J^+(S)\cap J^-(q)$. 
Then either $\overline{S\cap J^-(q)}$ is non-compact or strong causality
is violated at $q\in M$ (or both cases can happen).
\label{mhosztalyozas}
\end{lemma}

\noindent {\it Proof.} By definition and construction $\gamma_C$ is a 
future-inextendible non-spacelike half-curve in $J^+(S)\cap J^-(q)$. 
Subsequently, by \cite[Lemma 8.2.1]{wal2} and Lemma \ref{kompaktsag} here we 
get the result. $\Diamond$
\vspace{0.1in}

\noindent Now we can turn our attention to the existence of physically
relevant examples of space-times possessing the Malament--Hogarth property. 
Lemma \ref{mhosztalyozas} indicates that the class of
Malament--Hogarth space-times can be divided into two major subclasses:
the first one contains space-times in which an infinite, non-compact
portion of a spacelike submanifold is visible from some event. There is an 
abundance of such examples: any maximal extension of the 
Reissner--Nordstr\"om, Kerr \cite{ete-nem}, (universal cover of 
the) anti-de Sitter \cite{hog1, hog2} are examples.

The second subclass consists of those which lack strong causality along 
the future Cauchy horizon of their maximal extension. A maximally extended 
Taub--NUT space-time, certain extendible Gowdy space-times possess this 
property however the corresponding inextendible curves are incomplete i.e., 
have finite lengths only.\footnote{Hereby we acknowledge that \cite[Proposition 
2.5]{ete} is false because it is based on an erroneous calculation.} In 
other words these non-globally hyperbolic space-times are generalized 
Malament--Hogarth space-times only. At this moment we cannot answer the 
question whether or not this second subclass of Malament--Hogarth 
space-times is empty.

After getting some feeling of Malament--Hogarth space-times we
indicate their relationship with the strong cosmic censorship scenario. 
The content of Lemma \ref{mhosztalyozas} is that the Malament--Hogarth
property implies the Geroch--Horowitz--Penrose property for non-globally
hyperbolic space-times. However, the converse is not necessarily true as 
we have seen. But the converse seems to be true at least for asymptotically 
flat or hyperbolic space-times. Guided by these observations we cannot 
resist the temptation to exhibit a sharper formulation of the strong 
cosmic censor conjecture as follows \cite{ete}.

\begin{conjecture} {\em (sharpening of the Geroch--Horowitz--Penrose
version of the strong cosmic censor conjecture)} Let $(S,h,k)$ be an
asymptotically flat or asymptotically
hyperbolic (i.e., anti-de Sitter) initial data for Einstein's equation
(this implies $(S,h)$ is geodesically complete). Suppose a fundamental
matter field is given represented by its stress-energy tensor $T$
satisfying the dominant energy condition. 

Then, if the maximal Cauchy development of this
initial data set is extendible, this extension is a Malament--Hogarth
space-time and Malament--Hogarth events lie on or beyond the future Cauchy 
horizon $H^+(S)$ in the extension.
\label{sejtes}
\end{conjecture}

\noindent In analogy with the refomulated Theorem \ref{GHP'} this 
conjecture may be refomulated as well.

\begin{conjecture} {\em (reformulation of the sharpening)} Let $(S,h,k)$
be an initial data set for Einstein's equation as in Conjecture
\ref{sejtes}. Then if the maximal Cauchy development of this space-time
is extendible, for each $x\in H^+(S)$ in any extension, $J^+(S)\cap
J^-(x)$ contains a future-directed timelike curve of infinite length.
\label{sejtes'}
\end{conjecture}

\noindent A promising attack on this conjecture, straightforward by this
reformulation is to study the so-called ``radiation problem'' formulated
in the introduction of \cite{chr-o'm}. The authors address the problem
of finding points whose causal futures are complete in the Cauchy
development of a given asymptotically flat initial data.

In light of our considerations sofar Conjecture \ref{sejtes'} can be read 
such a way that a non-globally hyperbolic asymptotically flat or anti-de 
Sitter space-time contains a gravitational computer capable to break the 
Turing barrier. Therefore the problem of the existence of such space-times 
is apparently the same as that of computers capable of performing non-Turing 
computations. 


\section{Concluding remarks: what has been proved?}
\label{four}


One conclusion of our considerations here is that the problem of the {\it 
strong cosmic censorship} naturally splits up into two parts: suppressing 
the problem of genericity or stability in the formulation we obtain a causal or 
conformal variant {\bf SCCC-GHP} in Section \ref{one} (i.e., Theorems 
\ref{GHP} and \ref{GHP'}) whose proof is easy (essentially a 
consequence of the definition of global hyperbolicity). Meanwhile putting the 
emphasis onto the genericity or stability of extendible space-times we obtain 
a more geometric version like {\bf SCCC} in Section \ref{one} and run into the 
well-known technical difficulties. Unlike {\bf SCCC-GHP}, the {\bf SCCC} is 
``very much open''. But we have learned that {\bf SCCC-GHP} (i.e.,
Theorems \ref{GHP} and \ref{GHP'}) has an appropriate geometric modification 
namely {\bf SCCC-MH} in Section \ref{one} (i.e., Conjectures \ref{sejtes} 
and \ref{sejtes'}). 

The other conclusion is that in dealing with the usual formulation 
{\bf SCCC} of the strong cosmic censor conjecture one also seems to 
encounter (through the concept of a gravitational computer and {\bf SCCC-MH}) 
certain very general variants of the Church--Turing thesis namely 
{\bf Ph-ChT} in Section \ref{one} controlling the computational 
capacity of a broad class of physical computers (called gravitational 
computers here). This indicates that the strong cosmic censor conjecture in 
its full depth might be not only technically but even 
conceptually an extraordinary difficult problem. 

However we have to emphasize again that our speculations require future 
work: for example it is important to understand if other asymptotically 
flat or hyperbolic, extendible space-times admit the Malament--Hogarth 
property or not. It would be also interesting to know if the aforementioned 
new type of Malament--Hogarth space-times (i.e., which violate the strong 
causality along their Cauchy horizons) exist or not. 

Neverthless if our considerations turn out to be correct then we can 
establish an intimate link betwix the strong cosmic censor conjecture, a 
problem situated in the heart of recent theoretical physics and
computability theory, a subject previously considered as a pure
mathematical discipline.
\vspace{0.1in}

\noindent {\bf Acknowledgement.} The author is grateful to I. R\'acz, 
E. Minguzzi and L.B. Szabados for the stimulating discussions. The work was 
partially supported by OTKA grant No. NK81203 (Hungary).


\begin{thebibliography}{99}

\bibitem{acq-gas} Acquaviva, G., Gaspar, Y.: {\it Gravity and 
complexity}, Eur. Phys. Journ. Plus {\bf 127}: 65 (2012);

\bibitem{ale-ion-kla} Alexakis, S., Ionescu, A.D., Klainerman, S.: {\it
Uniqueness of smooth stationary black holes in vacuum: small perturbations
of the Kerr spaces}, preprint, arXiv: {\tt 0904.0982 [gr-qc]},
39pp (2009);

\bibitem{bee-ehr-eas} Beem, J.K., Ehrlich, P.E., Easley, K.L.: {\sl Global
Lorentzian geometry}, Second Edition, Marcel Dekker, Inc. New York (1996);

\bibitem{bla} Blaha, S.: {\it A quantum computer foundation for the
standard model and superstring theories}, preprint, , ArXiv: {\tt 
hep-th/0201092}, 78pp (2002);

\bibitem{bra} Bray, H.L.: {\it Proof of the Riemannian Penrose
inequality using the positive mass theorem}, Journ. Diff. Geom. {\bf 59},
177-268 (2001);

\bibitem{bri-hor} Brill, D.R., Horowitz, G.T.: {\it Testing cosmic
censorship with black hole collisions}, Phys. Rev. {\bf D49}, 840-852
(1994);

\bibitem{cal-pav} Calude, C.S., Pavlov, B.: {\it Coins, quantum
measurements and Turing's barrier}, Quantum Information Processing {\bf 1}, 
107--127 (2002);

\bibitem{chr1} Christodoulou, D.: {\it Examples of naked singularity
formation in gravitational collapse of a scalar field}, Ann. Math. {\bf
104}, 607-665 (1994);

\bibitem{chr2} Christodoulou, D.: {\it The instability of naked
singularities in the gravitational collapse of a scalar field}, Ann. Math.
{\bf 149}, 183-217 (1999);

\bibitem{chr-o'm} Christodoulou, D., O'Murchadha, N.: {\it The boost 
problem in general relativity}, Comm. Math. Phys. {\bf 80}, 271-300 (1981);

\bibitem{chru} Chru\'sciel, P.T.: {\it On uniqueness in the large of
solutions of Einstein equations (``strong cosmic censorship'')},
Australian National Univ. Press, Canberra (1991);

\bibitem{chr-ise-mon} Chru\'sciel, P.T., Isenberg, J., Moncrief, V.:
{\it Strong cosmic censorship in polarized Gowdy spacetimes}, Class.
Quant. Grav. {\bf 7}, 1671-1680 (1990);

\bibitem{chr-ren} Chru\'sciel, P.T., Rendall, A.D.: {\it Strong cosmic
censorship in vacuum space-times with compact, locally homogeneous
Cauchy surfaces}, Ann. Phys. {\bf 242}, 349-385 (1995);

\bibitem{daf1} Dafermos, M.: {\it Stability and instability of the Cauchy 
horizon for the spherically-symmetric Einstein--Maxwell-scalar field 
equations}, Ann. Math. {\bf 158}, 875-928 (2003);

\bibitem{daf2} Dafermos, M.: {\it The interior of charged black holes and 
the problem of uniqueness in general relativity}, Comm. Pure Appl. Math. 
{\bf 58}, 445-504 (2005);

\bibitem{ear} Earman, J.: {\sl Bangs, crunches, whimpers and shrieks}, 
Oxford Univ. Press, Oxford (1995);

\bibitem{ete} Etesi, G.: {\it Note on a reformulation of the strong 
cosmic censor conjecture based on computability}, Phys. Lett. {\bf
B550}, 1-7 (2002);

\bibitem{ete-nem} Etesi, G., N\'emeti, I.: {\it Non-Turing
computations via Malament--Hogarth space-times}, Int. Journ. Theor.
Phys. {\bf 41}, 341-370 (2002);

\bibitem{fre} Freedman, M.H.: {\it P/NP, and the quantum field computer}, 
Proc. Natl. Acad. Sci. USA {\bf 95}, 98-101 (1998);

\bibitem{ger-har} Geroch, R.P., Hartle, J.: {\it Computability and 
physical theories}, Found. Phys. {\bf 16}, 533-550 (1986);

\bibitem{ger-hor} Geroch, R.P., Horowitz, G.T.: {\it Global structure of
space-time}, in: {\sl General relativity, an Einstein centenary survey}, ed.:
Hawking, S.W., Israel, W., Cambridge Univ. Press, Cambridge (1979); 

\bibitem{haw-ell} Hawking, S.W., Ellis, G.F.R.: {\sl The large scale
structure of space-time}, Cambridge Univ. Press, Cambridge (1973);

\bibitem{her-hor-mae} Hertog, T., Horowitz, G.T., Maeda, K.: {\it
Generic cosmic censorship violation in anti-de Sitter space},
Phys. Rev. Lett. {\bf 92}, 131101 (2004);

\bibitem{hog1} Hogarth, M.L.: {\it Non-Turing computers and non-Turing
computability}, in: East Lansing: Philosophy of Science Association
{\bf 1}, ed.: Hull, D., Forbens, M., Burian, R.M., 126-138 PSA (1994);  

\bibitem{hog2} Hogarth, M.L.: {\it Deciding arithmetic using SAD computers}, 
British Journ. Phil. Sci. {\bf 55}, 681-691 (2004);

\bibitem{hol} Holzegel, G.: {\it A note on the instability of
Lorentzian Taub--NUT space}, Class. Quant. Grav. {\bf 23}, 3951-3962
(2006);

\bibitem{hor-she} Horowitz, G.T., Sheinblatt, H.J.: {\it Tests of cosmic
censorship in the Ernst spacetime}, Phys. Rev. {\bf D55}, 650-657 (1997);

\bibitem{hui-ilm} Huisken, G., Ilmanen, T.: {\it The Riemannian Penrose
inequality}, Int. Math. Res. Not. {\bf 20}, 1045-1058 (1997);   

\bibitem{kie1} Kieu, T.D.: {\it Quantum algorithm for Hilbert's tenth
problem}, Int. Journ. Theor. Phys. {\bf 42}, 1461-1478 (2003);

\bibitem{kie2} Kieu, T.D.: {\it A reformulation of Hilbert's tenth problem 
through quantum mechanics}, Proc. Roy. Soc. {\bf A460}, 1535 (2004);

\bibitem{kla-rod-sze} Klainerman, S., Rodnianski, I., Szeftel, J.: {\it
Overview of the proof of the bounded $L^2$ curvature conjecture},
preprint, arXiv: {\tt 1204.1772 [math.AP]}, 133pp (2012);

\bibitem{min} Minguzzi, E.: {\it Limit curve theorems in Lorentzian geometry}, 
Journ. Math. Phys. {\bf 49}, 092501 (2008);
 
\bibitem{ori1} Ori, A.: {\it The structure of the singularity inside a
realistic rotating black hole}, Phys. Rev. Lett. {\bf 68}, 2117-2120 (1992);

\bibitem{pen1} Penrose, R.: {\it Gravitational collapse: the role of
general relativity}, Rev. Nuovo Cim. {\bf 1}, 252-276 (1969);

\bibitem{pen2} Penrose, R.: {\it Naked singularities}, Ann. New
York Acad. Sci. {\bf 224}, 125-134 (1973);

\bibitem{pen3} Penrose, R.: {\it Singularities and time asymmetry}, in:
{\sl General relativity, an Einstein centenary survery}, ed.:
Hawking, S.W., Israel, W., Cambridge Univ. Press, Cambridge (1979);

\bibitem{pen4} Penrose, R.: {\it The question of cosmic censorship},
Journ. Astrophys. Astr. {\bf 20}, 233-248 (1999);

\bibitem{poi-isr} Poisson, E., Israel, W.: {\it Internal structure of
black holes}, Phys. Rev. {\bf D41}, 1796-1809 (1990);

\bibitem{rac1} R\'acz, I.: {\it On the existence of Killing vector 
fields}, Class. Quant. Grav. {\bf 16}, 1695-1703 (1999);

\bibitem{rac2} R\'acz, I.: {\it Symmetries of spacetimes and their 
relation to initial value problems}, Class. Quant. Grav. {\bf 18}, 
5103-5113 (2001);

\bibitem{rin} Ringstr\"om, H.: {\it Cosmic censorship for Gowdy
spacetimes}, Living Rev. Relativity {\bf 13}, 2 (2010); 

\bibitem{sch} Schlesinger, K-G.: {\it On the universality of string
theory}, Found. Phys. Lett. {\bf 15}, 523-536 (2002);

\bibitem{wal1} Wald, R.M.: {\it Gedanken experiment to destroy a black
hole}, Ann. Phys. {\bf 82}, 548-556 (1974);

\bibitem{wal2} Wald, R.M.: {\sl General relativity}, Univ. Chicago Press, 
Chicago (1984);

\bibitem{wel} Welch, P.D.: {\it The extent of computation in
Malament--Hogarth space-times}, British Journ. Phil. Sci. {\bf 59},
659-674 (2008).


\end{thebibliography}
\end{document}